%% file: main.tex
\documentclass[journal]{IEEEtran}

\usepackage{graphicx}
\ifCLASSINFOpdf
\else
\fi
\usepackage{url}
\usepackage{amsfonts}
\usepackage[noadjust]{cite}

\usepackage{xcolor}
\usepackage{subcaption}

\usepackage{algorithm}
\usepackage{algpseudocode}
\usepackage{amsmath}
\usepackage{amssymb}
\usepackage{amsthm}
\usepackage{soul}
\usepackage{booktabs}
\usepackage{multirow}

\usepackage{subfiles}

\title{The Implications of Decentralization in Blockchained Federated Learning: Evaluating the Impact of Model Staleness and Inconsistencies}

\author{
\IEEEauthorblockN{Francesc Wilhelmi$^{\star}$, Nima Afraz$^{\sharp}$, Elia Guerra$^{\flat}$, and Paolo Dini$^{\flat}$\vspace{0.1cm}
}\\
\IEEEauthorblockA{$^{\star}$\emph{Radio Systems Research, Nokia Bell Labs, Stuttgart, Germany}}\\
\IEEEauthorblockA{$^{\sharp}$\emph{CONNECT Centre, School of Computer Science, University College Dublin, Dublin, Ireland}}\\
\IEEEauthorblockA{$^{\flat}$\emph{Sustainable Artificial Intelligence, Centre Tecnològic de Telecomunicacions de Catalunya, Barcelona, Spain}}
\IEEEauthorblockN{\thanks{Corresponding author: \emph{francisco.wilhelmi@nokia.com}.}
\thanks{This work has been partially funded by the Spanish project PID2020-113832RB-C22(ORIGIN)/MCIN/AEI/10.13039/50110001103 and by FREE6G - TSI-063000-2021-151 from the Ministerio de Asuntos Económicos y Transformación Digital and the European Union – NextGenerationEU under the framework of the “Plan de Recuperación, Transformación y Resiliencia” and the “Mecanismo de Recuperación y Resiliencia”.}}
}
\begin{document}

\maketitle

\begin{abstract}
Blockchain promises to enhance distributed machine learning (ML) approaches such as federated learning (FL) by providing further decentralization, security, immutability, and trust, which are key properties for enabling collaborative intelligence in next-generation applications. Nonetheless, the intrinsic decentralized operation of peer-to-peer (P2P) blockchain nodes leads to an uncharted setting for FL, whereby the concepts of FL round and global model become meaningless, as devices' synchronization is lost without the figure of a central orchestrating server. In this paper, we study the practical implications of outsourcing the orchestration of FL to a democratic setting such as in a blockchain. In particular, we focus on the effects that model staleness and inconsistencies, endorsed by blockchains' modus operandi, have on the training procedure held by FL devices asynchronously. Using simulation, we evaluate the blockchained FL operation \textcolor{black}{by applying two different ML models (ranging from low to high complexity) on the well-known MNIST and CIFAR-10 datasets, respectively,} and focus on the accuracy and timeliness of the solutions. Our results show the high impact of model inconsistencies on the accuracy of the models (up to a ~35\% decrease in prediction accuracy), which underscores the importance of properly designing blockchain systems based on the characteristics of the underlying FL application.
\end{abstract}


\IEEEpeerreviewmaketitle

\subfile{sections/1_introduction}

\subfile{sections/2_related_work}

\subfile{sections/3_background}

\subfile{sections/4_problem_description}

\subfile{sections/5_experimental_results}

\subfile{sections/6_future_directions}

\subfile{sections/7_conclusions}

\ifCLASSOPTIONcaptionsoff
\newpage
\fi
	
\bibliographystyle{IEEEtran}
\bibliography{bibliography}

\end{document}

%% file: sections/1_introduction.tex
\section{Introduction}
\label{sec:introduction}

\subsection{The decentralization of machine learning}

Enabled by the advances in edge computing, the decentralization of artificial intelligence (AI) and machine learning (ML) unlocks a prominent paradigm where intelligence is brought closer to end-users. Decentralized AI allows real-time and near real-time applications to meet the ever-increasing latency requirements~\cite{zhou2019edge} thanks to the delay reduction achieved by eliminating the communication with a central node performing model inference. And not only that, but decentralized AI can potentially save energy by reducing the burden of massive data transferring, and also distributing the workload across many sites, rather than hosting it in a data center~\cite{guerra2023cost}.

To enable edge intelligence and address scalability issues in ML, federated learning (FL) emerged in 2016 as a powerful tool to train ML models in a distributed manner~\cite{konevcny2016federated}. In FL, a set of participants (also referred to as \textit{FL clients} or \textit{FL devices}) train an ML model collaboratively by exchanging model parameters, rather than by exchanging training data explicitly. Following this approach, an FL algorithm---see, e.g., Federated Averaging (FedAvg)~\cite{mcmahan2017communication}---can potentially reduce the overheads of ML training and also enhance the privacy of its centralized counterpart. FL was initially defined around the operation of a central server, which is responsible for orchestrating the ML training procedure by iteratively retrieving ML model updates from FL clients, computing global ML model updates, and distributing the outputs back to FL devices. Given the potential weaknesses of the centralized setting~\cite{li2020federated}, including security, bottlenecking, or straggling issues, alternative decentralized architectures for FL have been recently proposed~\cite{pokhrel2020decentralized}. 

To enable the decentralization of FL, blockchain technology~\cite{nakamoto2008bitcoin} stands as an appealing approach, as blockchains provide secure, immutable, and trustworthy decentralized storage. The decentralized realization of FL through blockchain, referred to as blockchained FL (sometimes it is also referred to as \textit{FLchain}~\cite{majeed2019flchain}), provides trust via cryptographic proof to federated ecosystems where multiple (often unreliable) parties cooperate to train a shared model. The blockchained FL framework does not only address centralization issues (e.g., single point of failure) but also provides complementary mechanisms that may boost FL settings, including effective ways of incentivizing FL participants (e.g., through native tokenization) to undertake ML model training~\cite{wang2020porx}.

\subsection{Challenges of blockchained federated learning}
\label{section:challenges_blockchain}

\begin{figure*}[t!]
\includegraphics[width=\linewidth]{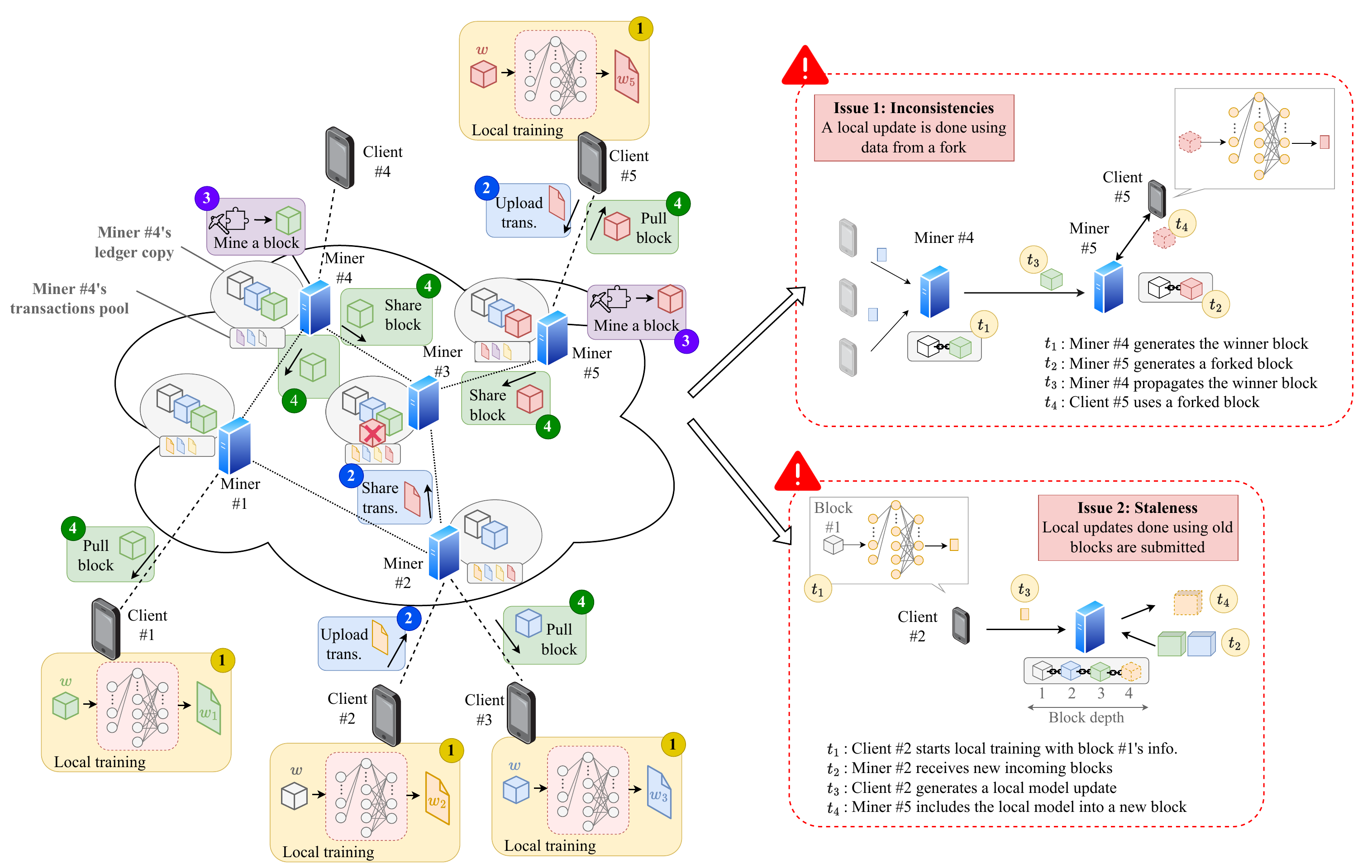}
\caption{Overview of the procedures carried out by FL devices and blockchain nodes in blockchained FL.}
\label{fig:flchain}
\end{figure*}

Although blockchain enables security and trustworthiness in decentralized FL applications, its inherent decentralization entails some important implications that must not be disregarded. Figure~\ref{fig:flchain} summarizes the blockchained FL operation by graphically showing the following main procedures (highlighted by the numbered circles in the figure):
\begin{enumerate}
    \item \textbf{On-device computation at end devices:} FL clients generate local ML model updates by training a model using their local data. The resulting local models are encapsulated in blockchain transactions.
    \item \textbf{Exchange of transactions:} The local model updates are submitted by FL devices to the blockchain. Blockchain nodes (e.g., miners) maintain a shared pool of transactions synchronized by exchanging and broadcasting transactions through peer-to-peer (P2P) messages.
    \item \textbf{Blockchain writing:} Blockchain miners take transactions from the shared pool to generate new blocks that update the status of the ledger. Each block corresponds to a newly computed aggregate model, built from FL client model updates. 
    \item \textbf{Block propagation and consensus:} Blockchain nodes exchange blocks and enforce consensus rules to ensure that the ledger is consistent across the entire P2P network. The latest blocks are provided to FL devices to continue training the FL model iteratively.
\end{enumerate}

Under ideal conditions in which information is instantaneously propagated without errors, the blockchain would keep track of the overall FL optimization process as in centralized FL, thus storing a global update in each block (equivalent to performing FL rounds). However, such ideal conditions are unfeasible and do not hold in reality. First, the delays associated with information propagation in the blockchain, together with the decentralized nature of consensus, make the ledger inconsistent in different parts of the blockchain network. Apart from that, FL devices have heterogeneous capabilities in terms of computation and communication, thus contributing to the inconsistent usage of models for training.

In this work, we focus our attention on the two following issues derived from the operation of blockchained FL:
\begin{itemize}
    \item \textbf{Issue \#1 - Ledger inconsistencies:} A ledger inconsistency (e.g., a fork in the main chain) arises as a result of the concurrent and decentralized mining operation. The effects of ledger inconsistencies on the FL operation are exemplified in the right-top part of Fig.~\ref{fig:flchain}. In the provided example, Miner \#4 generates a valid block at $t_1$, but before such a block is propagated at $t_3$, Miner \#5 generates another valid block at $t_2$ (notice that $t_3 > t_2$) with potentially different model updates. Until Miner \#5's chain is correctly updated (i.e., after Miner \#4's block is reaffirmed by another mined block), the forked block (shown in red) is used by Client \#5 to generate a new local update at time $t_4$, thus leading to a model inconsistency.
    \item \textbf{Issue \#2 - Model staleness:} The delays for computing and securing local updates on the blockchain lead FL devices to potentially use \textcolor{black}{outdated models} for model training\textcolor{black}{, i.e., models that are trained insufficiently compared to newer client updates but which are still to be processed by the blockchain}. Model staleness may raise concerns in specific datasets and scenarios, provided that old model updates can negatively impact the global model's accuracy. As an example of model staleness, in the right-bottom part of Fig.~\ref{fig:flchain}, Client \#2 uses the model in Block \#1 to generate a new local update at $t_1$. However, before the update from Client \#2 is computed and secured on the blockchain (which entails computation, communication, and mining delays), newer blocks with fresher models are generated as a result of the asynchronous operation of other FL clients. Then, the question is whether the stale model update generated by Client \#2 is still valid or not.
\end{itemize}

As described above, the blockchained FL setting leads to a set of issues that can potentially affect the learning procedure therein. To the best of our knowledge, these issues have been barely studied in the literature. 

\subsection{Our contributions}

This paper is an extended version of our previous work presented in~\cite{wilhelmi2023decentralization}. It builds upon the same concepts and methodology to provide further characterization and insights into the blockchained FL setting and its inherent issues (highlighted in Section~\ref{section:challenges_blockchain}). The specific contributions of this paper are as follows:
\begin{itemize}
    \item We provide a comprehensive and self-contained overview of blockchained FL, including its technological realization and the description of the issues associated with it.
    \item We characterize the main implications that arise from blockchained FL, i.e., ledger inconsistencies and model staleness. In this regard, we compute the freshness of blocks and use it as a staleness metric in FL.
    \item We provide a simulation tool that is new of its kind for realistically capturing the blockchained FL operation. This tool, which is named Block\textit{FL}sim~\cite{wilhelmi2022flchainsim}, integrates BlockSim~\cite{alharby2019blocksim} with Pytorch~\cite{NEURIPS2019_9015}, thus allowing to simulate blockchained FL applications.
    \item We evaluate the blockchained FL approach through extensive simulations and assess the impact of ledger inconsistencies and model staleness on FL accuracy. Our evaluation is carried out for automatic image recognition in different scenarios. \textcolor{black}{More specifically, we consider two different sub-problems, characterized by the MNIST~\cite{deng2012mnist} and the CIFAR-10~\cite{krizhevsky2009learning} datasets for the evaluation of blockchained FL.}
\end{itemize}

\subsection{Structure of the document}

The remainder of this paper is structured as follows. Section~\ref{section:related_work} provides an overview of the state-of-the-art solutions regarding decentralized realizations of FL and the integration of blockchain and FL. Section~\ref{section:background} describes background concepts on blockchain and FL as independent solutions and then delves into the blockchained FL paradigm. The system model is presented in Section~\ref{section:system_model}, which is followed by the \textcolor{black}{performance evaluation using simulations} in Section~\ref{section:experimental_results}. \textcolor{black}{Section~\ref{section:future_directions} provides insights on future research directions} and Section~\ref{section:conclusions} concludes the paper with final remarks.

%% file: sections/2_related_work.tex
\section{Related Work}
\label{section:related_work}

\subsection{Decentralized federated learning}

The decentralization of ML procedures has been popularized in the past years as a result of the increased computational and storage capabilities of handheld devices and the increasing reluctance to share private data with a server. A proposal to distribute in a P2P fashion the well-known stochastic gradient descend (SGD) algorithm, normally used for training purposes, is presented in~\cite{pmlr-v119-koloskova20a}. In~\cite{lalitha2018fully}, a fully decentralized mechanism was proposed to train ML models by leveraging one-hop communications between neighbor nodes. Similarly, gossip-based communications~\cite{ormandi2013gossip} were leveraged in~\cite{miozzo2021distributed} to train ML models in a P2P network. As the authors of~\cite{miozzo2021distributed} showed, besides improving robustness, gossip learning significantly reduces energy consumption while leading to the same performance as in centralized learning.

The idea of gossip learning has also been applied in decentralized FL. In particular, the work in~\cite{hu2019decentralized} leveraged gossip communications to allow the exchange of model segments among workers, which were used to provide ML model updates in a decentralized manner. A decentralized version of FedAvg whereby clients communicate with their neighbors was also proposed and analyzed in ~\cite{sun2022decentralized}. A similar approach was presented in~\cite{xing2020decentralized}, where decentralized FL was realized through device-to-device (D2D) communications. In addition, the authors in~\cite{xing2020decentralized} defined protocols for both analog and digital types of transmission modes, which were evaluated through simulations. Another decentralized FL solution was proposed in~\cite{qu2021decentralized} to accommodate the specific needs of unmanned aerial vehicles (UAV) networks. An alternative mechanism for decentralizing FL can be found in~\cite{roy2019braintorrent}, which was inspired by the BitTorrent protocol. More specifically, through the mechanism in~\cite{roy2019braintorrent}, FL clients can request model updates to other devices on-demand.

Between centralized and fully decentralized FL, other solutions relied on cloud-edge computing hierarchies~\cite{wang2019adaptive,liu2020client} or clustering capabilities~\cite{chen2022cfl} in order to improve the performance of FL. These classes of hybrid methods are useful to boost efficiency and mitigate the single point of failure and scalability issues of centralized FL, and at the same time prevent synchronization issues and model inconsistencies as in fully decentralized FL. For a more comprehensive overview of existing decentralized ML and FL solutions, we refer the interested reader to the surveys in~\cite{beltran2022decentralized, khan2023decentralized, gabrielli2023survey}.

\subsection{Blockchained federated learning}

Blockchained FL was first introduced in~\cite{kim2019blockchained, majeed2019flchain} as a prominent solution for decentralizing FL and replacing the figure of the centralized orchestrating server with a blockchain. Different types of solutions have been envisioned to realize blockchained FL~\cite{nguyen2021federated, hou2021systematic}. Important design considerations lie in where and how to deploy the blockchain, which largely impacts the cost and performance of the solution~\cite{afraz2023blockchain}. 

A common approach in the literature relies on mobile edge computing (MEC) servers, co-located within access points (APs) or base stations (BSs) connected to end-users, to perform blockchain operations such as transaction validation or mining. This approach is very convenient for fulfilling typical blockchain requirements, as edge servers are typically equipped with high computation and communication capabilities. In~\cite{nguyen2021federated}, for instance, a generic blockchained FL architecture based on edge computing was defined. Through this approach, FL devices (in charge of performing model training) can submit ML model updates to the closest AP/BS, where an edge server acts as a full blockchain node (it collects transactions and participates in the block mining). Other works following a similar approach can be found in~\cite{majeed2019flchain, lu2020blockchain, zhao2020privacy, wilhelmi2022analysis}, while alternatives including cloud/fog computing solutions were considered in \cite{qu2020decentralized, qu2021decentralized, ali2021integration, nguyen2020integration}.

Apart from the system's architecture, the election of the blockchain type is critical to the desired performance and capabilities and, therefore, should be influenced by the underlying FL application and the degree of trust among participants. In this regard, public blockchains using consensus mechanisms such as Proof of Work (PoW) have been adopted in~\cite{wilhelmi2022analysis, qu2021decentralized} to accommodate fully decentralized applications. Conversely, consortium and private blockchains using Algorand or Practical Byzantine Fault Tolerance (PBFT) have been considered in~\cite{zhao2020privacy, qi2021privacy} to fit more restricted settings where a limited group of participants maintain the blockchain.

When it comes to the evaluation of blockchained FL applications, analytical models like the one in~\cite{kim2019blockchained} have been widely adopted to derive the end-to-end latency model of a blockchained FL system. Similarly, end-to-end latency models were proposed in~\cite{lu2020low, pokhrel2020federated} to characterize the communication, computation, and consensus delays in blockchained FL. 

Analytical models like the abovementioned ones, while providing a good intuition of the performance of blockchained FL applications, typically assume unrealistic conditions like perfect synchronization or proper FL client scheduling (see, e.g.,~\cite{pokhrel2020decentralized}). In contrast, blockchained FL implementations suit better an asynchronous setting in which FL clients operate independently. The asynchronous blockchained FL problem has been studied in~\cite{liu2021blockchain,wang2022asynchronous, feng2022bafl} but, still, the impact that blockchain decentralization has on the FL operation has been little studied to date. For instance, in the literature, forks have been considered to affect the transaction confirmation time only (see, e.g.,~\cite{kim2019blockchained}), so their implications on the FL training procedure remain unclear. In this paper, we aim to cover this gap in the literature and study the issues associated with model staleness and ledger inconsistencies that naturally arise in practical blockchained FL realizations.

%% file: sections/3_background.tex
\section{Blockchain and Federated Learning: Preliminaries}
\label{section:background}

In this section, we first introduce blockchain and FL as separate technologies, and we then delve into their confluence, leading to the blockchained FL setting.

\subsection{Blockchain}

\begin{figure}[ht!]
\centering
\includegraphics[width=\columnwidth]{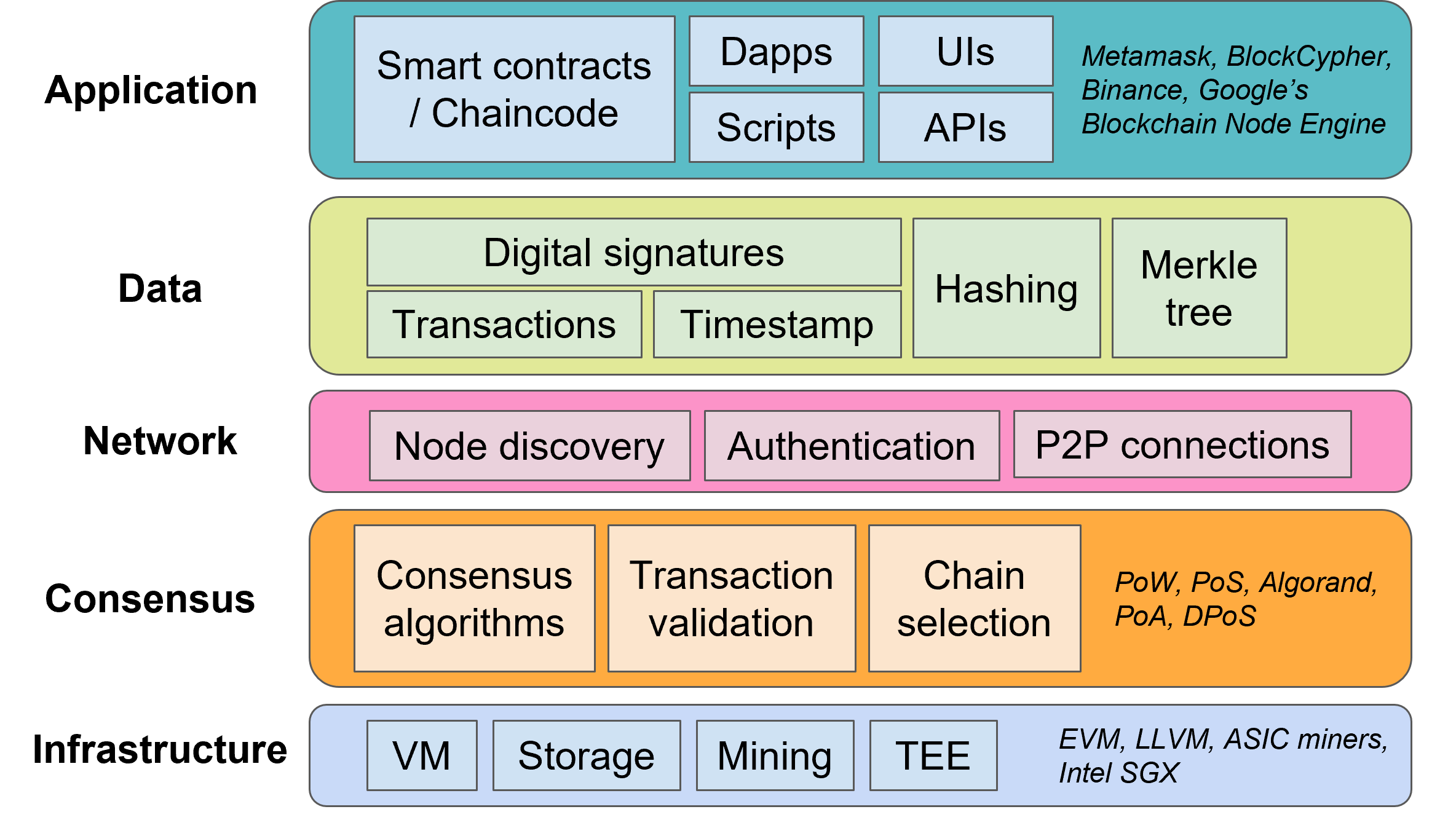}
\caption{Blockchain's protocol stack.}
\label{fig:blockchain_layers}
\end{figure} 

Blockchain is a distributed ledger technology where a set of P2P nodes holding their own copy of the ledger must agree on the history of timestamped transactions. Blockchain combines multiple concepts and technologies, including cryptography primitives, to achieve relevant properties such as security, privacy, and immutability in a decentralized data-sharing framework. To showcase the main components and operations of blockchains, we resort to the blockchain protocol stack illustrated in Fig.~\ref{fig:blockchain_layers} (other definitions have also been proposed in~\cite{kan2018multiple, tseng2020blockchain}), which includes the following layers:
\begin{itemize}
    \item \textbf{Application:} The application layer deals with the decentralized applications (e.g., finance, supply chain, \textcolor{black}{IoT~\cite{wang2020secure}}) running on the blockchain. This includes all the infrastructure and tools for generating and handling the decentralized application data (e.g., transactions, smart contracts) hosted in the blockchain and may include the support from user graphical interfaces, APIs, wallets, etc.
    \item \textbf{Data:} The data layer is related to the way data is structured and stored in the blockchain. This layer applies concepts such as the chaining of blocks or digital signatures to enforce privacy, security, immutability, and transparency properties.
    \item \textbf{Network:} The network layer allows blockchain nodes communicate to exchange transactions and blocks, and also to enforce consensus. In this layer, procedures like node discovery or block/transaction exchanges are defined.
    \item \textbf{Consensus:} The consensus layer establishes the rules to be applied by blockchain nodes in order to participate in the maintenance and update of the distributed ledger. Consensus mechanisms such as PoW establish the operations of blockchain miners for creating and accepting new blocks, thus being the core of the decentralized and concurrent synchronization of the ledger.
    \item \textbf{Infrastructure:} Finally, the infrastructure layer defines the set of physical devices, connections, and operations within the underlying blockchain network. Depending on the type of consensus adopted, varying computational and storage capabilities might be required.
\end{itemize}

\begin{figure}[ht!]
\centering
\includegraphics[width=\columnwidth]{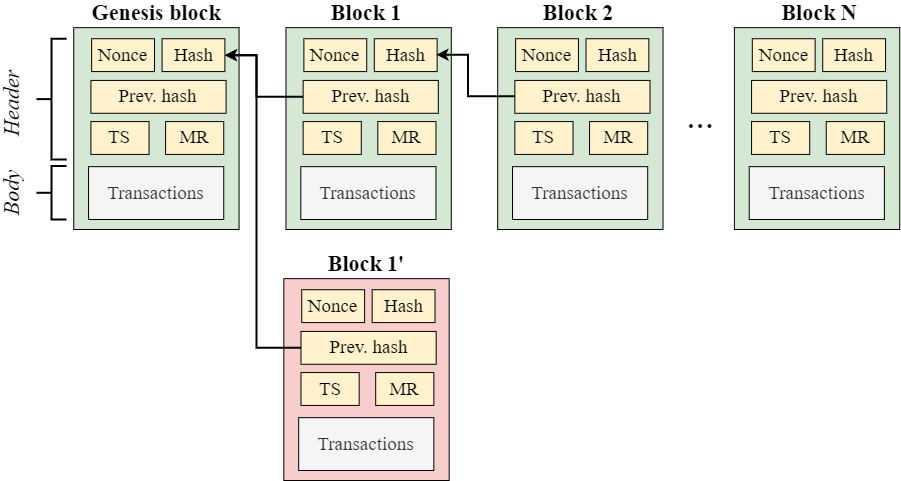}
\caption{Blockchain structure and block information.}
\label{fig:blockchain_structure}
\end{figure}

The data and the consensus layers constitute the core of blockchain technology, as they define the way information is structured and validated by participants, which is the key to providing security and immutability. As shown in Fig.~\ref{fig:blockchain_structure}, transactions (i.e., events updating the ledger) are grouped in blocks, each chained after the previous one and starting by the \textit{genesis block} (with depth 0). Depending on the blockchain realization (e.g., Bitcoin, Ethereum, Hyperledger Fabric), blocks may carry different types of information, but in general, blocks include the following basic fields:
\begin{itemize}
    \item \textbf{Nonce (\textit{number only used once}):} The number used to demonstrate a cryptographic proof within the PoW mining operation. In Bitcoin, a nonce is a 32-bit number that, when hashed with the rest of the headers of a block, meets the difficulty imposed by the consensus protocol (i.e., the solution of such a hash contains a predefined amount of initial zeros).
    \item \textbf{Hash:} Output of the hash function (e.g., SHA-256) when combining the headers of the block. The hash is used as cryptographic proof to authorize a given miner to add a new block to the ledger.
    \item \textbf{Timestamp (TS):} The time the block was created.
    \item \textbf{Merkle root (MR):} Typically, for building the hash of a block, transactions are organized in a Merkle tree to further guarantee immutability. The root of the Merkle tree is included in the header.
    \item \textbf{Body of transactions:} Set of transactions generated by the application layer and which are included in the block by the  miner.
\end{itemize}

Blockchain consensus plays a major role in preserving the integrity of the data stored in a blockchain and it is required for blockchain participants to adopt the same history of the ledger. Particularly in permissionless blockchains,\footnote{In \textit{permissionless} blockchains, different from \textit{permissioned} ones, any party can participate in the consensus procedure.} where miners work concurrently, mining can lead to forks, i.e., different (forked) versions of the ledger adopted by different blockchain nodes. In PoW, which has been widely used in public blockchains (e.g., Bitcoin, Ethereum), forks occur when two or more miners come up with a valid nonce before the winning miner shares its block with the rest of the miners. Forks are solved by enforcing consensus rules, such as adopting the longest chain (with more invested power) as the valid one. Following the consensus rules, a blockchain node in a forked chain eventually switches to the main chain, which occurs when the main chain obtains more confirmations than any other forked version.\footnote{In Bitcoin, for instance, up to six confirmations are required before a transaction (a payment) is considered to be secure.}

\subsection{Federated learning}

In FL~\cite{aledhari2020federated}, a set of clients collaborate to train a global model in a distributed manner. To that purpose, instead of sharing raw data directly (as done in traditional centralized ML applications), FL devices sequentially exchange model parameters, obtained by training the latest received global model on local data. Broadly speaking, the goal of FL is to minimize a global finite-sum cost function, weighted by the contribution of each node (e.g., in terms of data samples with respect to the total length of the distributed dataset).

\begin{figure}[ht!]
\centering
\includegraphics[width=.85\columnwidth]{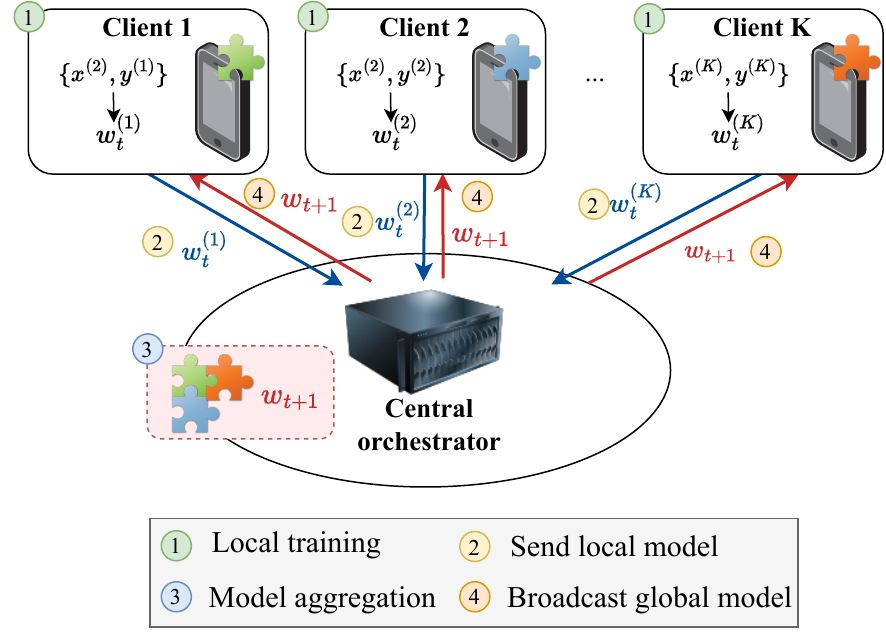}
\caption{ML model training in Federated Averaging (FedAvg).}
\label{fig:fedavg}
\end{figure}

FedAvg (illustrated in Fig.~\ref{fig:fedavg} and described in Algorithm~\ref{alg:fedavg}) is one of the most popular algorithms to carry out the FL operation~\cite{mcmahan2017communication}. In FedAvg, the parameter server (or central orchestrating server) selects a subset $S_t\subseteq \mathcal{K}$ of FL clients in each round $t$. The selected clients compute a local update $w_{t}^{(k)}$ from the current global model $w_t$ by running SGD. Then, the server aggregates the received local models from the selected clients to generate a new global model update $w_{t+1}$, which is sent to the next set of selected clients for further training.

\begin{algorithm}[ht!]
	\caption{Federated Averaging (FedAvg) }\label{alg:fedavg}
	\begin{algorithmic}[1]
	    \State \textbf{Initialize:} Initial model $w_0$, batch size $B$, number of epochs $E$, learning rate $\eta$
		\For{$t=0,\ldots,T$}
		\State Select $S_t \subseteq \mathcal{K}$
		\For{$k \in S_{t}$}
		\State Pull $\boldsymbol{w}_{t}$ from central server: $\boldsymbol{w}^{(k)}_{t,0}=\boldsymbol{w}_{t}$
		\For{$e=1,\ldots,E$}
		\State Update local model: $\boldsymbol{w}^{(k)}_{t,e}=\boldsymbol{w}_{t,e}^{(k)}-\eta\nabla l^{(k)}_{t,e}$
		\EndFor
		\State Push $\boldsymbol{w}_{t+1}^{(k)} \leftarrow \boldsymbol{w}_{t,E}^{(k)}$
		\EndFor
		\State $\boldsymbol{w}_{t+1}=\frac{1}{ \vert S_t \vert}\sum_{k\in S_t} \boldsymbol{w}_{t+1}^{(k)}$
		\EndFor
	\end{algorithmic}
\end{algorithm}

\subsection{Blockchained federated learning}
\label{sec:blockchained_fl}

The blockchained FL solution relies on a distributed ledger to securely store ML model updates from FL clients. This solution gets rid of the central orchestrating server and, instead, empowers a decentralized network of miners to maintain the status of the FL global model by updating the ledger based on FL clients' updates. The ledger can be accessed asynchronously by the FL clients for either reading (e.g., downloading the latest global models) or writing (e.g., submitting fresh local model updates). Under such an asynchronous and uncoordinated setting where each FL client works independently and maintains its own FL model history, the concepts of FL training round and global model lose meaning.

\begin{figure}[ht!]
\includegraphics[width=.9\columnwidth]{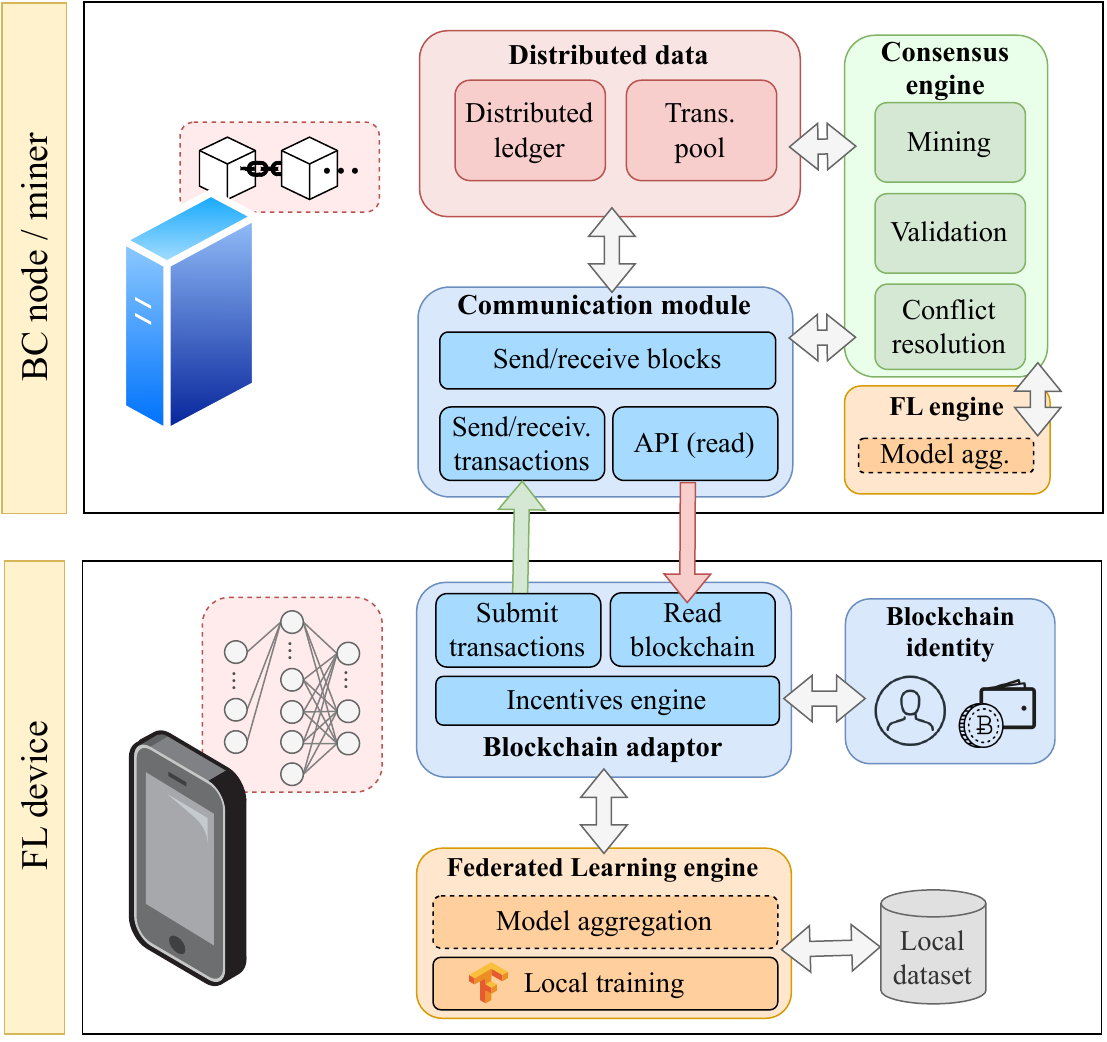}
\caption{Blockchained FL logical components and interactions.}
\label{fig:flchain_components}
\end{figure}

To carry out decentralized FL through blockchain, we identify two types of logical entities, namely \textit{blockchain nodes} and \textit{FL clients}, each one with specific functionalities and logical components (illustrated in Fig.~\ref{fig:flchain_components}). Blockchain nodes are responsible for gathering, sharing, and verifying transactions and maintaining the distributed ledger, while FL clients perform a given federated optimization task by contributing with locally trained models. The set of functional blocks from blockchain nodes and FL clients is as follows: 
\begin{itemize}
    \item \textbf{Distributed data module:} Used to store and keep track of the history of blocks and to maintain the pool of unconfirmed transactions.
    \item \textbf{Consensus engine:} Deals with the enforcement of the distributed protocol through mining operations (e.g., PoW, PBFT), validation, and conflict resolution.
    \item \textbf{Communication module:} It allows blockchain nodes to share either transactions or blocks among them. The communication module also provides an interface with FL clients for gathering transactions (local models) and exposing the ledger.
    \item \textbf{Blockchain adaptor:} It allows clients to interact with the blockchain to either submitting transactions or retrieving the latest FL models. An incentive engine can be used to encourage the participation of FL clients in the distributed learning operation.
    \item \textbf{Federated learning engine:} Allows training a model using on-device ML libraries (e.g., TensorFlow) and local data (it can be an offline process). In this paper, model aggregation is considered to be done by miners. Nevertheless, the envisioned architecture also allows performing model aggregation on the FL device side, as previously proposed in~\cite{kim2019blockchained, nguyen2021federated}.
\end{itemize}

The specific implementation of blockchain nodes and FL clients depends on the underlying FL application and the requirements therein. Based on the type of blockchain adopted (e.g., public vs. private, permissioned vs. permissionless blockchains), blockchain nodes and miners might need different hardware capabilities, thus leading to either specialized or general-purpose devices meeting certain storage and computation requirements. In public permissionless blockchains, blockchain nodes must possess high computational and storage capabilities to support computation-intensive mining and store a large history of transactions, respectively. In Bitcoin, for instance, blockchain nodes are typically specialized devices (e.g., ASIC miners) with tens to hundreds of terahash per second (TH/s) power and hundreds of GBs of memory~\cite{taylor2017evolution}. When it comes to FL clients, they need enough storage to keep their local dataset (typically in the order of a few GB, depending on the application) and from low to moderate computational power to perform local training. FL clients are typically end devices like smartphones or laptops, but other solutions leveraging MEC-based computation offloading exist~\cite{liu2020client}.

%% file: sections/4_problem_description.tex
\section{System Model}
\label{section:system_model}

\subsection{Blockchained FL model}

A set of $\mathcal{K}=\{1,2,...,K\}$ clients collaborate to train a global model $w$ by sharing locally trained model parameters $w^{(k)}\in \mathbb{R}^d$ through a blockchain. Starting from a global model provided by block $b$, $w^{(b)}$, each device $k$ attempts to minimize a local loss function $l^{(k)}(\cdot)$ by running $E$ epochs of SGD on its local data. Using a dataset $\mathcal{D}^{(k)}$ with $D^{(k)}=|\mathcal{D}^{(k)}|$ samples, a client $k$ updates the local model parameters $w^{(k)}$ as:
\begin{equation}
    w^{(k)} = w^{(b)} - \eta_l \nabla l^{(k)}(w^{(b)},\mathcal{D}^{(k)}),
\label{eq:2}
\end{equation}
where $\eta$ is the learning rate and $\nabla l^{(k)}(w^{(b)},\mathcal{D}^{(k)})$ is the $k$-th client average loss gradient with respect to $w^{(b)}$. A client $k$ counts on $\rho^{(k)}$ computational power for training a local model.

The local model updates are aggregated and the resulting global model is included into a blockchain block. The aggregation of local updates, assumed to be done by the miner generating the block $b$, is computed as
\begin{equation}
    w^{(b)} = \sum_{k \in b} \frac{D^{(k)}}{D^{(b)}} w^{(k)},
\end{equation}
where $D^{(b)}=\sum_{k \in b} D^{(k)}$. Following this approach, any client can retrieve the latest block $b$ from its closest miner $m$ and use the latest global model to continue training the federated model.

\subsection{Blockchain latency model}

We consider a P2P network of $\mathcal{M}$ blockchain nodes acting as miners. The blockchain nodes are responsible for gathering transactions from FL devices, mining blocks, and broadcasting new transactions and blocks. The overall process, as described in the literature~\cite{kim2019blockchained,wilhelmi2022end}, can be decomposed into different steps, including transaction/block propagation, block mining, and consensus resolution, which delay characterizations are described in the following sections.

\subsubsection{Transaction and block propagation}
\label{section:communication_model}

Both transaction and block propagation delays are characterized by an exponential distribution with mean $1/\text{T}$, which is determined by the size of the data $\text{L}$ to be transmitted and the capacity $\text{C}_\text{link}$ of the link used. In particular, the mean transaction propagation latency is computed as
\begin{equation}
    \text{T}_\text{tp} = \frac{\text{L}_\text{t}}{\text{C}_\text{link}},
\end{equation}
where the transaction length $\text{L}_\text{t}$ is defined by the size of the ML model, which is computed as the number of model parameters ($\text{N}_\text{model}$) multiplied by 4 bytes (we assume that each model parameters is represented by a \textit{float32} variable).

Likewise, the mean block propagation delay is computed as
\begin{equation}
     \text{T}_\text{bp} = \frac{\text{L}_\text{b}}{\text{C}_\text{link}} = \frac{\text{L}_\text{bh} + \text{N}_\text{t}\cdot \text{L}_\text{t}}{\text{C}_\text{link}},
\end{equation}
where $\text{L}_\text{bh}$ is the block's header length and $\text{N}_\text{t}$ is the number of transactions carried in the block. In this paper, $\text{N}_\text{t}$ is fixed to 1, provided that miners are assumed to perform model aggregation and include the resulting global model in the block.

\subsubsection{Blockchain mining and consensus}
\label{section:blockchain_model}

We adopt a PoW-based type of consensus, whereby miners compete to update the blockchain by appening new blocks. In particular, each miner $m\in \mathcal{M}$ employs its computational hash power $\xi^{(m)}$ to generate a new block each time a valid block is received. The time it takes a miner $m$ to generate a block, $\text{T}_\text{bg}^{(m)}$, is characterized by an exponential distribution $\text{Exp}(\lambda^{(m)})$, being $\lambda^{(m)}$ defined as 
\begin{equation}
    \lambda^{(m)}= \frac{\xi^{(m)}}{\sum_{n\in\mathcal{M}} \xi^{(n)}} \frac{1}{BI},
\end{equation}
where $BI$ is the block interval, which, together with the total hash power ($\xi = \sum_{m\in \mathcal{M}} \xi^{(m)}$), determines the mining difficulty. 

As previously described, the mining operation associated with PoW is done concurrently and in a decentralized manner, which might lead to ledger inconsistencies in the form of forks. Assuming that the time between blocks is characterized by a Poisson inter-arrival process, the fork probability is given by
\begin{equation}
\begin{split}
    \text{P}_\text{fork} &= 1 - \prod_{\forall i \neq w} \Pr(\text{T}_\text{bg}^{(i)} - \text{T}_\text{bg}^{(w)} > \text{T}_\text{bp}^{(w)}) \\&= 1 - e^{-\mu (|\mathcal{M}|-1)\text{T}_\text{bp}^{(w)}},
\end{split}
\label{eq:fork_probability}
\end{equation}
where $\text{T}_\text{bg}^{(w)}$ and $\text{T}_\text{bp}^{(w)}$ are the winner's block generation and propagation delays, respectively. 

In PoW, forks are eventually solved by consensus, which states that the longest chain (i.e., the one with the highest invested computational power) is the valid one. However, from the point of view of a miner, consensus is not enforced until the next mined block is received, which allows switching to the version of the ledger accepted by the majority. The consensus resolution process, of course, has an impact on the models used by the FL devices for training, as the models used might differ depending on the miner providing its version of the ledger.

\subsection{Performance metrics}

To evaluate the performance of different blockchained FL realizations, we focus on the blockchain throughput and the model accuracy and staleness, defined in the following subsections.

\subsubsection{Blockchain throughput [transactions per second, TPS]}

The blockchain throughput is measured as the number of processed transactions per second. In particular, the effective throughput considers the transactions from the main chain only:
\begin{equation}
    \Gamma = \frac{\sum_{b\in \mathcal{B}_\text{main}}\text{N}_\text{t}^{(b)}}{\text{T}_\text{sim, total}},
\end{equation}
where $\mathcal{B}_\text{main}$ is the set of blocks in the main chain and $\text{T}_\text{sim, total}$ is the total simulated time.

\subsubsection{Model accuracy [percentage, \%]}

To evaluate the performance of the employed ML models, we use the classification accuracy, defined as
\begin{equation}
    \text{A} = \frac{\text{N}_\text{pred, correct}}{\text{N}_\text{pred, total}},
\end{equation}
where $\text{N}_\text{pred, correct}$ is the number of correct predictions and $\text{N}_\text{pred, total}$ is the total number of predictions done. We differentiate between two types of accuracy metrics, based on the data partition on which they are applied:
\begin{enumerate}
    \item Test accuracy, $\text{A}_\text{test}$: the accuracy measured in the complete test dataset once the training is finished. The model from the last mined block of the main chain is used for the evaluation.
    \item Block training/validation accuracy, $\text{A}_\text{train/val}^{(b)}$: the accuracy 
    achieved by the global model stored in a given block $b$ from the main chain. The training/validation data of the clients contributing to such a block is used for the measurement.
\end{enumerate}

\subsubsection{Model staleness [seconds, s]}

To measure the degree of staleness of the FL models stored in the blockchain, we resort to the blockchain and communication latency models provided in sections~\ref{section:blockchain_model} and~\ref{section:communication_model}. The freshness of a global model update does not only depend on the blockchain delays (including block mining and block propagation) but also on the delays associated with training local models by FL devices. In particular, the staleness of block $b$ is
\begin{equation}
    \Lambda^{(b)} = \frac{1}{|\mathcal{N}_\text{t}^{(b)}|} \sum_{n=1}^{|\mathcal{N}_\text{t}^{(b)}|} \big( \text{TS}_\text{bm}^{(b)} - \text{TS}_\text{tg}^{(b,n)} \big),
\end{equation}
where $\text{TS}_\text{tg}^{(b,n)}$ is the time at which transaction $n$ is generated, $\text{TS}_\text{bm}^{(b)}$ is the time block $b$ is mined, and $\mathcal{N}^{(b)}_\text{t}$ is the set of transactions used to generate the model in block $b$.

%% file: sections/5_experimental_results.tex
\section{\textcolor{black}{Performance Evaluation}}
\label{section:experimental_results}

To carry out the experiments, we use Block\textit{FL}sim~\cite{wilhelmi2022flchainsim}, an extension of BlockSim~\cite{alharby2019blocksim} that provides event-based simulations of blockchained FL applications. Block\textit{FL}sim uses Pytorch libraries~\cite{NEURIPS2019_9015} to train and evaluate FL models following the scheme previously illustrated in Fig.~\ref{fig:flchain}. Through the simulation of blockchained FL, we can characterize the phenomena associated with model staleness and inconsistencies, as discussed in Section~\ref{sec:introduction}. The simulation parameters used for the considered scenarios are collected in Table~\ref{table:simulation_parameters}.

\begin{table}[ht!]
\centering
\caption{Simulation parameters.}
\label{table:simulation_parameters}
\begin{tabular}{@{}ccc@{}}
\toprule
Parameter & Description & Value \\ \midrule
$BI$ & Block interval & \{1, 10, 60\} s \\
$\text{N}_\text{t}$ & Max. local models per block & \{1, 5, 10\}\\
$\text{L}_\text{t}$ & Transaction length (\textcolor{black}{MNIST}/CIFAR-10) & \textcolor{black}{0.796}/2.327 Mb \\
$\text{L}_\text{bh}$ & Block header length & 20 Kb \\ 
$M$ & Number of miners & 10 \\
$\text{C}_\text{link}$ & P2P links' capacity & \{1, 100\} Mbps \\
$\text{N}_\text{b}$ & Total simulated blocks & 200 \\
\hline
$K$ & Number of FL devices & \{10, 50, 100\} \\
$E$ & Number of local epochs & 5 \\
$B$ & Batch size & 64 \\
$\xi_\text{client}$ & Devices comp. power & 900 MIPS
\\
\bottomrule
\end{tabular}
\end{table}

\textcolor{black}{The evaluation is done on the MNIST~\cite{deng2012mnist} and CIFAR-10~\cite{krizhevsky2009learning} datasets, respectively. MNIST contains 70.000 samples ($28\times 28\times 1$ black and white images of hand-written numbers), split into 60.000 for training and 10.000 for test, and CIFAR-10 contains a total of 60.000 samples ($32\times 32\times 3$ color images of objects), split into 50.000 for training and 10.000 for test. In our case, for each dataset, we have further split the test partitions into test (30\%) and validation (70\%). The ML model selected to be trained in a federated manner is a Feed-forward Neural Network (FNN) for MNIST and a Convolutional Neural Network (CNN) for CIFAR-10. In both cases, cross-entropy is used as a loss function, which suits the target classification task well. The details of the implemented FNN and CNN are collected in Table~\ref{table:cnn_details}.}

\begin{table}[t!]
\centering
\caption{\textcolor{black}{Per-layer detailed information of the FNN-MNIST and CNN-CIFAR10 implementations.}}
\label{table:cnn_details}
\resizebox{\columnwidth}{!}{%
\begin{tabular}{@{}ccccccc@{}}
\toprule
\textbf{Model} & \textbf{Layer} & \textbf{Activation} & \textbf{Kernel} & \textbf{Stride} & \textbf{Input} & \multicolumn{1}{c}{\textbf{Output}} \\ \midrule
\multicolumn{1}{c}{\multirow{4}{*}{\textcolor{black}{FNN$_\text{MNIST}$}}} & \textcolor{black}{Fully-conn.} & \textcolor{black}{ReLU} & \textcolor{black}{-} & \textcolor{black}{-} & \textcolor{black}{784} & \textcolor{black}{200} \\
\multicolumn{1}{c}{} & \textcolor{black}{Fully-conn.} & \textcolor{black}{ReLU} & \textcolor{black}{-} & \textcolor{black}{-} & \textcolor{black}{200} & \textcolor{black}{200} \\
\multicolumn{1}{c}{} & \textcolor{black}{Fully-conn.} & \textcolor{black}{LogSoftMax} & \textcolor{black}{-} & \textcolor{black}{-} & \textcolor{black}{200} & \textcolor{black}{10} \\
\cmidrule(l){2-7} 
& \multicolumn{6}{c}{\textcolor{black}{Optimizer: SGD, learning rate = $10^{-3}$}} \\
\cmidrule(l){1-7} 
\multirow{9}{*}{CNN$_\text{CIFAR10}$} & Conv2D & ReLU & 3 & 1 & 3 & 16 \\
 & MaxPool & - & 2 & - & - & - \\
 & Conv2D & ReLU & 3 & 1 & 16 & 32 \\
 & MaxPool & - & 2 & - & - & - \\
 & Conv2D & ReLU & 3 & 1 & 32 & 64 \\
 & MaxPool & - & 2 & - & - & - \\
 & Fuly-conn. & ReLU & - & - & 1024 & 512 \\
 & Fully-conn. & ReLU & - & - & 512 & 64 \\
 & Fully-conn. & - & - & - & 64 & 10 \\ 
 \cmidrule(l){2-7} 
& \multicolumn{6}{c}{Optimizer: SGD, learning rate = $10^{-4}$, momentum = 0.9, weight decay = 5$\cdot10^{-4}$} \\
\cmidrule(l){1-7} 
\end{tabular}%
}
\end{table}

\subsection{Blockchain throughput}

\begin{figure}[ht!]
\centering
\includegraphics[width=0.8\columnwidth]{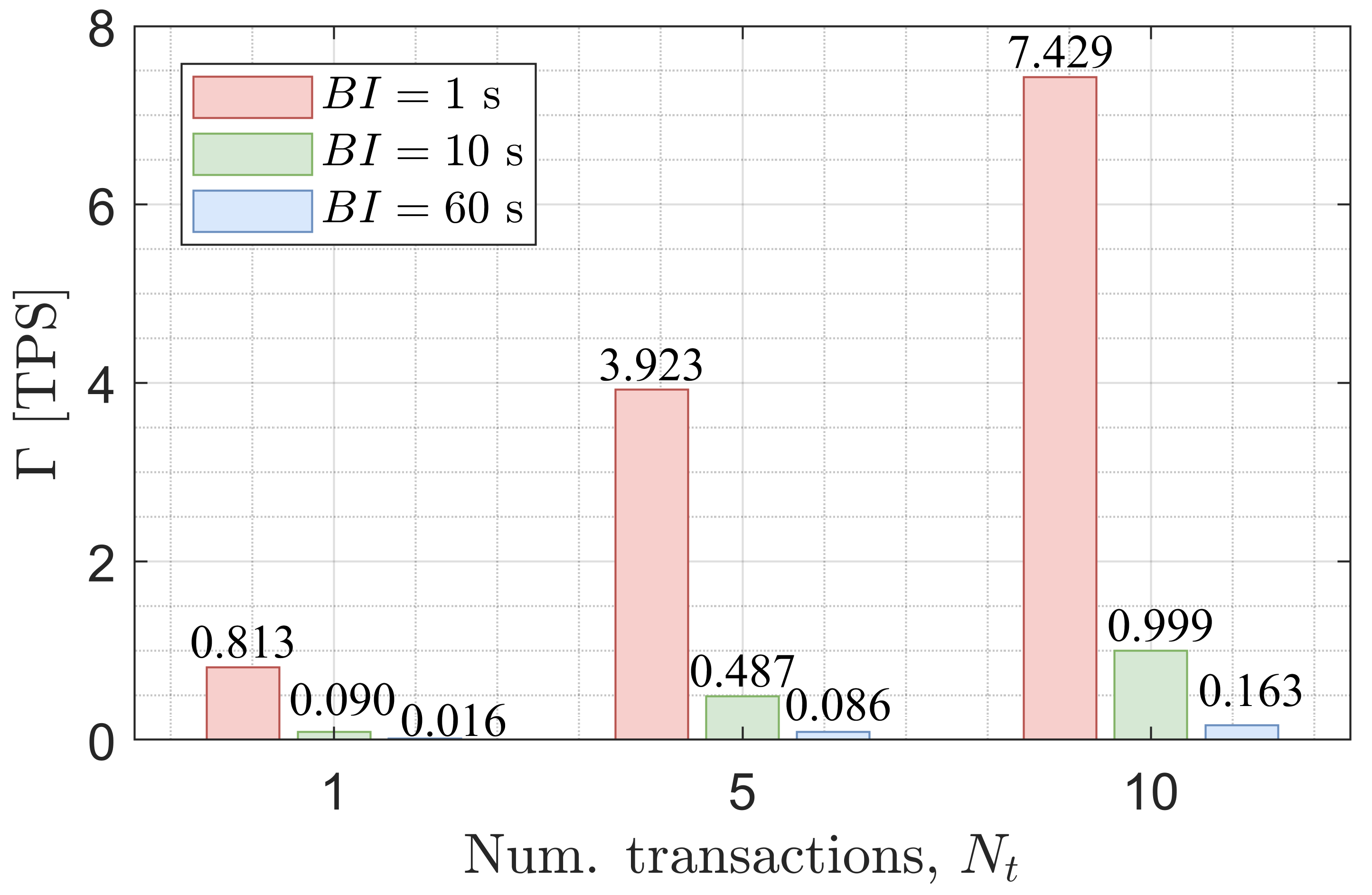}
\caption{Blockchain throughput achieved by each considered blockchain setting in TPS.}
\label{fig:blockchain_performance}
\end{figure}

First, we study the performance of different blockchain settings in Fig.~\ref{fig:blockchain_performance}, which shows the throughput in TPS achieved for different block interval ($BI$) values and numbers of transactions used per block ($\text{N}_\text{t}$). \textcolor{black}{To showcase the performance for the most challenging task considered, Fig.~\ref{fig:blockchain_performance} focuses on CIFAR-10, where a heavier CNN model is used.}

As shown, the blockchain throughput decreases as the block interval $BI$ increases (e.g., from 7.429 TPS to 0.163 for $BI=1$~s and $BI=60$~s, respectively, when $\text{N}_\text{t}=10$), which is mostly motivated by the bottleneck created by block mining. \textcolor{black}{Both $BI$ and $\text{N}_\text{t}$ are blockchain configuration parameters that allow for representing different types of blockchains (e.g., public, consortium, private). The block interval, for instance, is a fundamental aspect of the design of blockchains, provided that it determines the speed at which the blockchain validates and secures transactions.\footnote{In PoW mining, miners invest their computational power to prove the validity of the blocks they mine in a decentralized setting. Accordingly, large block intervals are derived from the necessity of enforcing a high mining difficulty in networks with a big computational power.} Typically, the block interval is selected according to the characteristics of the targeted scenario (e.g., type of participation access) and the security requirements of the supported decentralized application. Whereas there are multiple aspects involved in the establishment of security and trust in a blockchain, the block interval encapsulates well the characterization of different types of blockchain.} In particular, low block interval values such as  $BI=1$~s are feasible in blockchains where the miners are trusted (e.g., a private blockchain controlled by a consortium of network operators), while higher values like $BI=60$~s are required in blockchains where miners are trustless (e.g., a public blockchain where any interested party can participate). As for the set of included transactions used to generate each block, we observe that higher $\text{N}_\text{t}$ lead to higher throughput (e.g., from 0.813 to 7.429 TPS for $\text{N}_\text{t} = 1$ and $\text{N}_\text{t}=10$, respectively when $BI=1$~s), as blocks can carry more information. However, processing and validating a large volume of user transactions requires enough computational power from blockchain nodes. The blockchain throughput, as shown in more detail in the sequel, has an impact on the underlying application's performance, so it is an important metric to be optimized.

\subsection{FL model performance}

\begin{figure*}[ht!]
\centering
\begin{subfigure}{\columnwidth}
\includegraphics[width=\columnwidth]{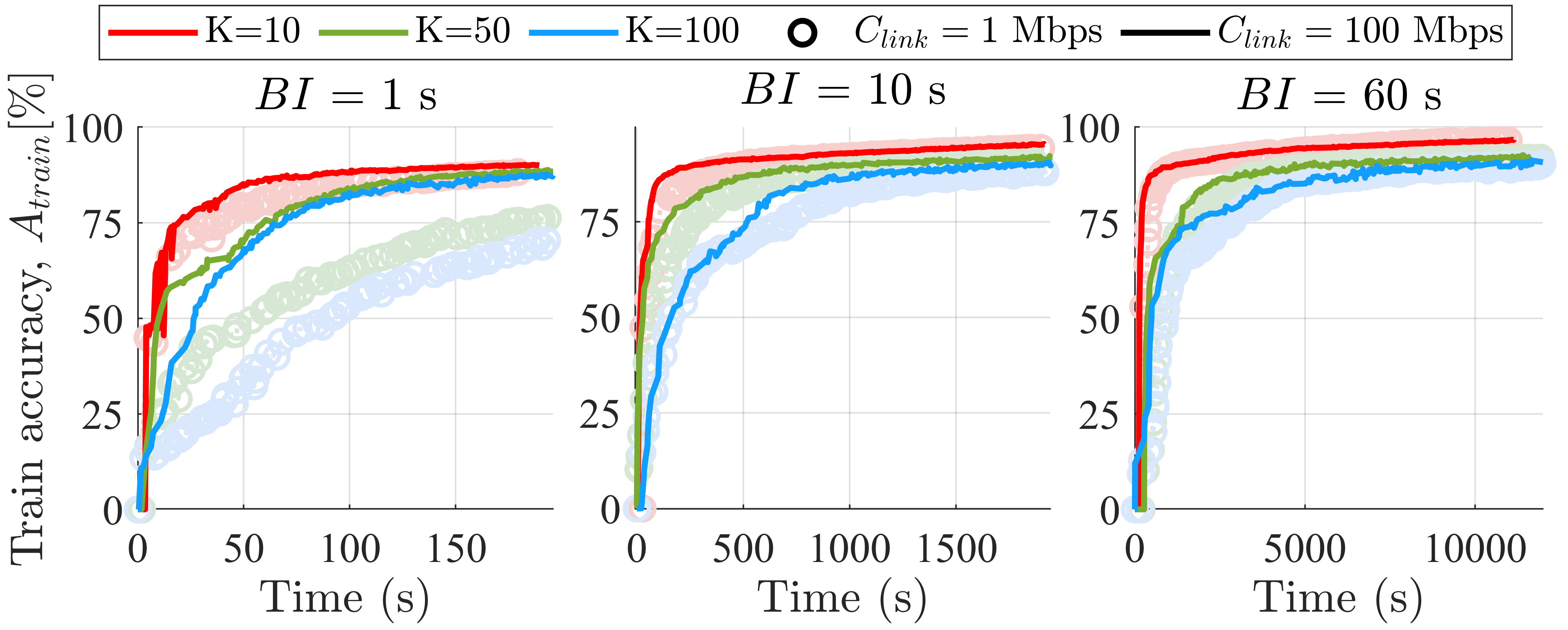}
\caption{\textcolor{black}{Training accuracy (MNIST).}}
\label{fig:training_accuracy_mnist}
\end{subfigure}
\hfill
\begin{subfigure}{\columnwidth}
\includegraphics[width=\columnwidth]{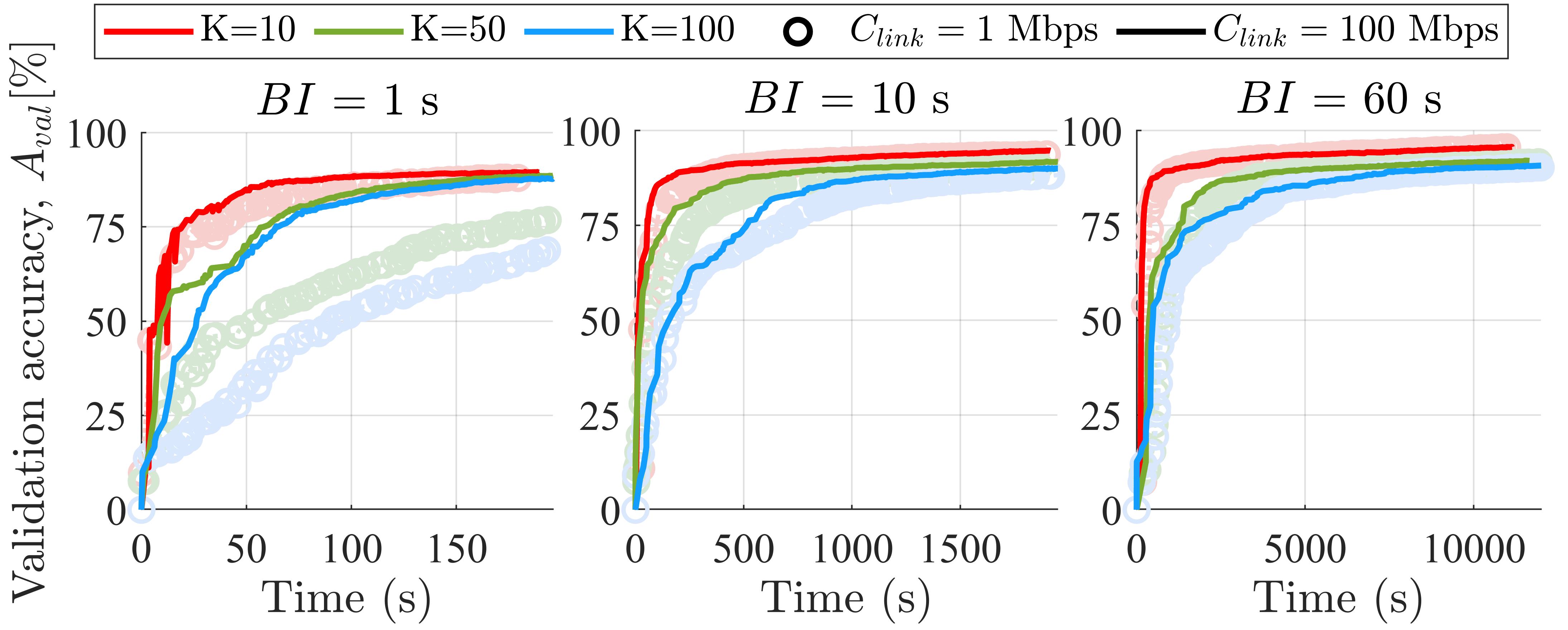}
\caption{\textcolor{black}{Validation accuracy (MNIST)}.}
\label{fig:validation_accuracy_mnist}
\end{subfigure}
\begin{subfigure}{\columnwidth}
\includegraphics[width=\columnwidth]{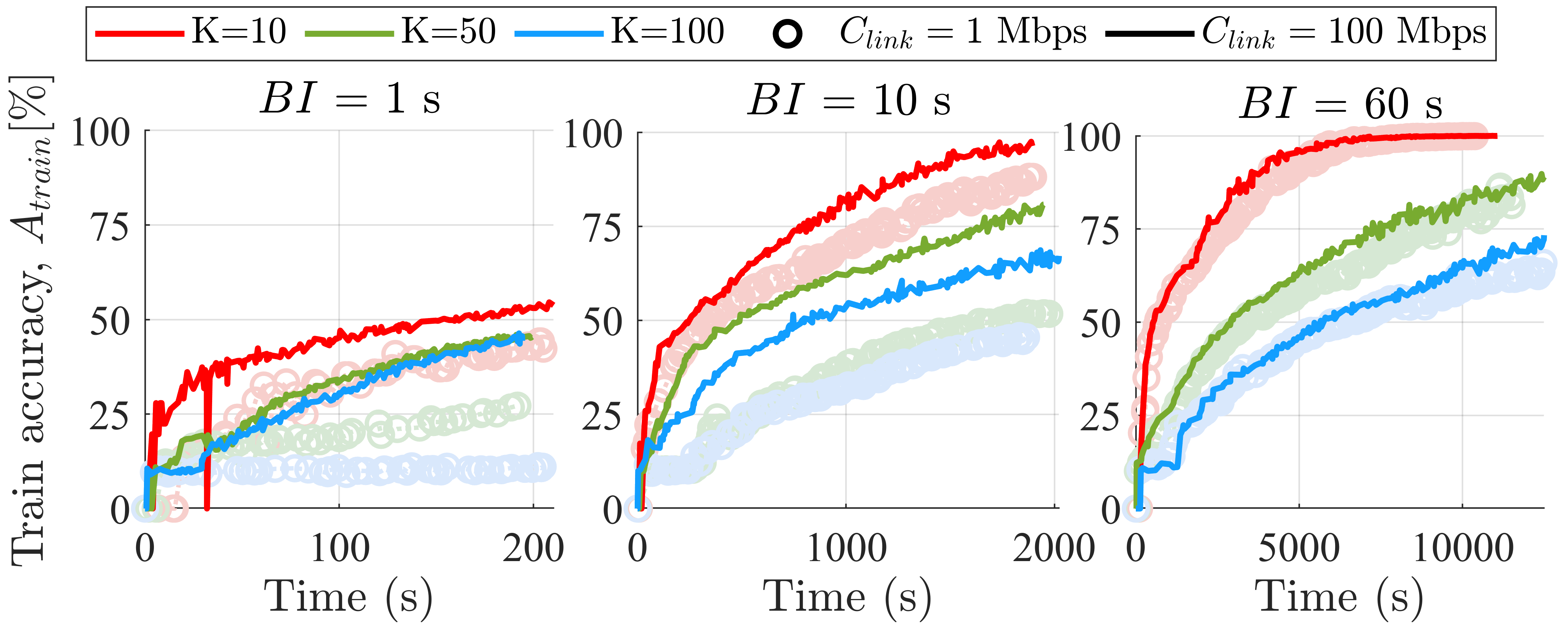}
\caption{\textcolor{black}{Training accuracy (CIFAR-10)}.}
\label{fig:training_accuracy}
\end{subfigure}
\hfill
\begin{subfigure}{\columnwidth}
\includegraphics[width=\columnwidth]{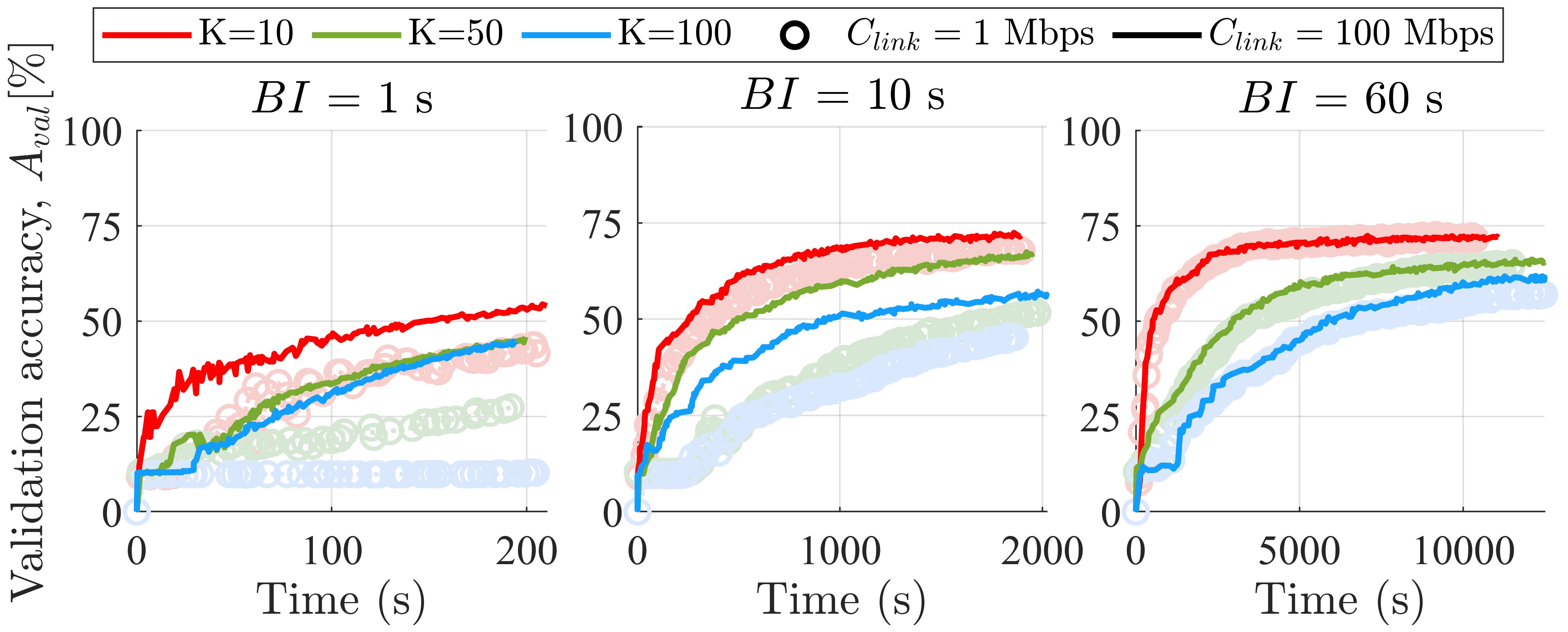}
\caption{\textcolor{black}{Validation accuracy (CIFAR-10).}}
\label{fig:validation_accuracy}
\end{subfigure}
\caption{Temporal evolution of the models' accuracy: \textcolor{black}{a) training accuracy (MNIST), b) validation accuracy (MNIST), c) training accuracy (CIFAR-10), d) validation accuracy (CIFAR-10)}. The experiments for $\text{C}_\text{link} = 1$~Mbps are represented by light circles, while $\text{C}_\text{link} = 100$~Mbps results are represented by solid lines.}
\label{fig:time_evolution_val_accuracy_main_chain}
\end{figure*}

\textcolor{black}{We now focus on the performance of the FL models in each blockchained FL setting. First, Fig.~\ref{fig:time_evolution_val_accuracy_main_chain} shows the temporal evolution of both the training (Fig.~\ref{fig:training_accuracy_mnist} and Fig.~\ref{fig:training_accuracy}) and the validation accuracy (Fig.~\ref{fig:validation_accuracy_mnist} and Fig.~\ref{fig:validation_accuracy}) achieved by the models carried in the blocks of the main chain, for the two different datasets.} The evaluation also includes the comparison of two types of P2P links, namely $\text{C}_\text{link} = \{1, 100\}$~Mbps. The maximum number of transactions per block, $\text{N}_\text{t}$, is set to 10.

\textcolor{black}{Starting with MNIST, we observe in Fig.~\ref{fig:training_accuracy_mnist} and Fig.~\ref{fig:validation_accuracy_mnist} that the performance is very similar for both training and validation accuracy, and this is due to the simplicity of the data employed (at which up to 99,7\% performance can be achieved~\cite{an2020ensemble}). For CIFAR-10 (see Fig.~\ref{fig:training_accuracy} and Fig.~\ref{fig:validation_accuracy}), in contrast, the discrepancies between training and validation accuracy are much more noticeable, being the training accuracy superior (up to 100\% accuracy) to the validation one (up to 75\%).}

\textcolor{black}{Regarding the different evaluation parameters, we find the following. First, the lower the block interval ($BI$), the worse the accuracy, especially in CIFAR-10, for which the model's complexity is higher. Setting $BI=1$~s, while allowing to generate blocks very fast, leads to a low training accuracy (in CIFAR-10, up to 50\% training/validation accuracy at the end of the training). As studied later with further detail, a short block generation time leads to a high number of forks, which is detrimental to the learning procedure. Increasing the block interval ($BI=10$~s and $BI=60$~s), instead, contributes to improving the accuracy significantly, as newly generated models are properly spread throughout the network, thus leading to a consistent federated training operation. Increasing the block interval, therefore, is one way of enforcing stability to FL training, but it leads to a higher training time.}

\textcolor{black}{When it comes to the number of FL participants ($K$), we observe that a higher performance is achieved as $K$ decreases in all the cases}. This is directly related to the amount of training data available to FL participants, provided that the entire dataset is split among all the considered participants. Apart from that, we observe that the gap in the accuracy achieved for different values of $K$ becomes bigger as $BI$ increases, as setting a higher $BI$ value allows for collecting more transactions in each block. In settings where $BI$ is small ($BI=1$~s), the probability of collecting enough transactions per block is low if the number of FL participants $K$ is low ($K=10$). In contrast, deployments with more FL clients in place ($K=100$) allow gathering plenty of local models per block, as there are more participants operating concurrently. However, the quality of the provided model updates for $K=100$ is lower than for $K=10$, as data is split among a higher number of users. Therefore, there is a trade-off between the number of local models per block and the quality of the same, and such a trade-off becomes apparent when the block interval is not properly dimensioned according to the underlying FL application.

Finally, regarding the type of links used to transmit transactions and blocks, we observe different behaviors depending on the blockchain setting. For low block interval values, i.e., $BI = \{1, 10\}$~s, there is a big gap between the accuracy achieved by $\text{C}_\text{link}=100$~Mbps (shown by the solid lines) and $\text{C}_\text{link} = 1$~Mbps (shown by the light circles). This gap is due in large part to the need for propagating the updated models throughout the blockchain network consistently, thus enabling a robust FL training operation. Otherwise, ledger inconsistencies lead to conflicting and redundant efforts in iterating the global model, thus negatively affecting the overall model accuracy. The gap between the performance achieved by each $\text{C}_\text{link}$ value, however, is mitigated as $BI$ increases, thanks to the model consistency achieved in those cases.

\subsection{Model inconsistencies}

\begin{figure}[ht!]
\centering
\includegraphics[width=\columnwidth]{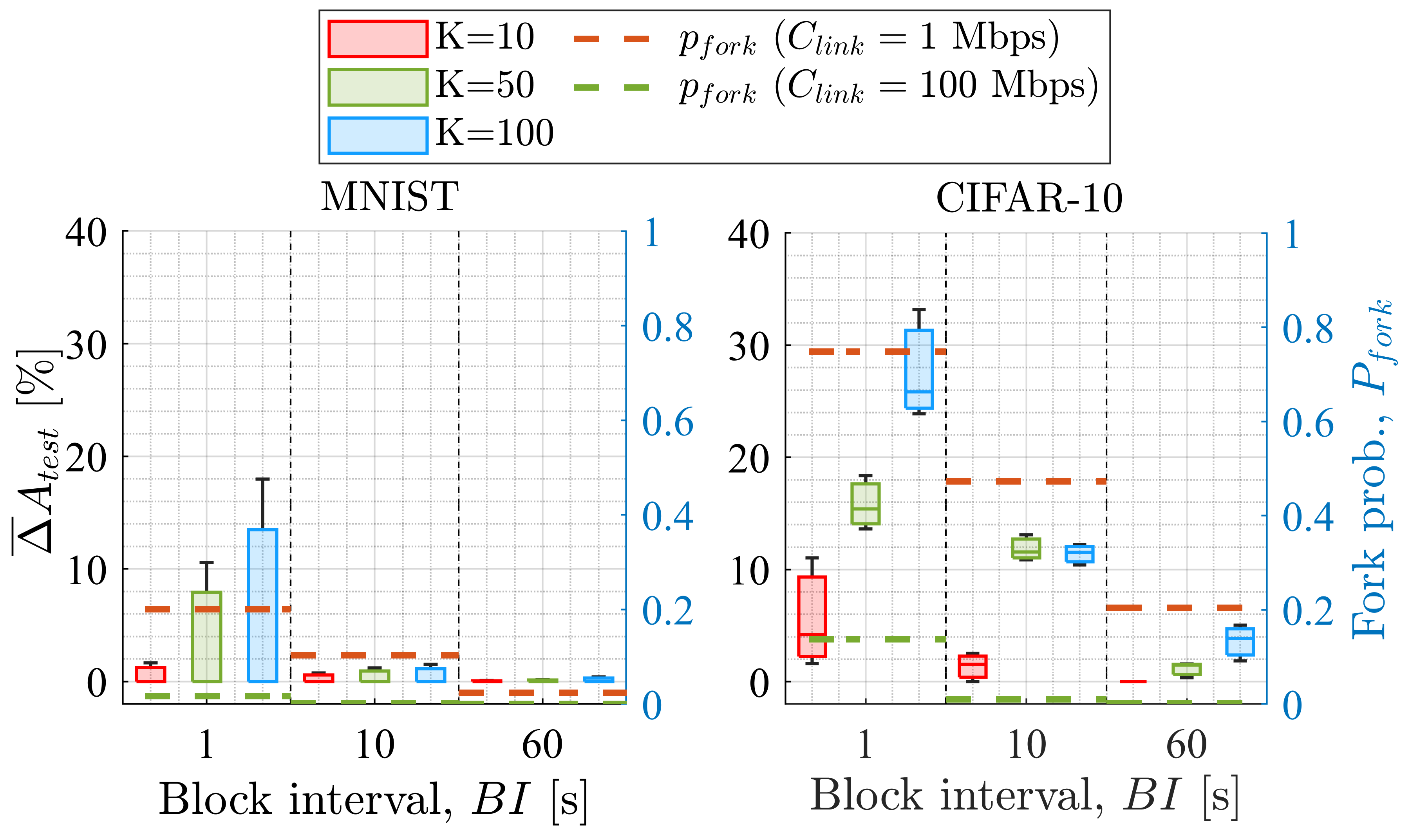}
\label{fig:difference_mnist}
\caption{Effect of forks on the test accuracy \textcolor{black}{for both MNIST (left plot) and CIFAR-10 (right plot)}. The boxplots show the mean difference in the accuracy achieved at various blockchained FL settings when using either $\text{C}_\text{link} = 1$~Mbps or $\text{C}_\text{link} = 100$~Mbps. The horizontal dashed lines represent the mean fork probability in each case.}
\label{fig:difference_test_accuracy_forks}
\end{figure}

As shown before, the type of blockchain adopted and its characteristics have a significant impact on the performance of the ML model, being the block interval ($BI$) and the P2P links capacity ($\text{C}_\text{link}$) some of the most critical parameters in that regard. Now, to further illustrate the impact of the blockchain characteristics on the federated models' accuracy, in Fig.~\ref{fig:difference_test_accuracy_forks}, we show the difference between the test accuracy achieved for $\text{C}_\text{link} = 1$~Mbps (leading to a higher fork probability) and $\text{C}_\text{link} = 100$~Mbps (granting more stability), accompanied by the fork probability in each case. The fork probability is a good indicator of the model inconsistencies, as it gives a good intuition of how many blockchain versions (i.e., global models) are concurrently used by different FL clients.

The results in Fig.~\ref{fig:difference_test_accuracy_forks} confirm the observations done above for Fig.~\ref{fig:time_evolution_val_accuracy_main_chain}, as the highest differences in the test accuracy are obtained for low block interval values ($BI = \{1, 10\}$~s) and high numbers of FL participants ($K = {50, 100\\}$ clients). In particular, a high fork probability ($\text{P}_\text{fork}\approx 0.8$ for the worst case in CIFAR-10) can lead to a decrease of up to 30\% on the model's accuracy when users' diversity is high, i.e., for $K=100$ clients. In these situations, increasing the pace at which information is processed within the blockchain by increasing $BI$ is helpful to minimize the fork probability, thus mitigating the effects of model inconsistencies. \textcolor{black}{For instance, for MNIST, $BI=10$~s provides a good trade-off between the block confirmation time and the fork probability. For CIFAR-10, which uses heavier modes, setting $BI=60$~s is required to keep the test accuracy difference below 5\%.}

\subsection{Model staleness}

\begin{figure}[ht!]
\centering
\begin{subfigure}{\columnwidth}
\includegraphics[width=\columnwidth]{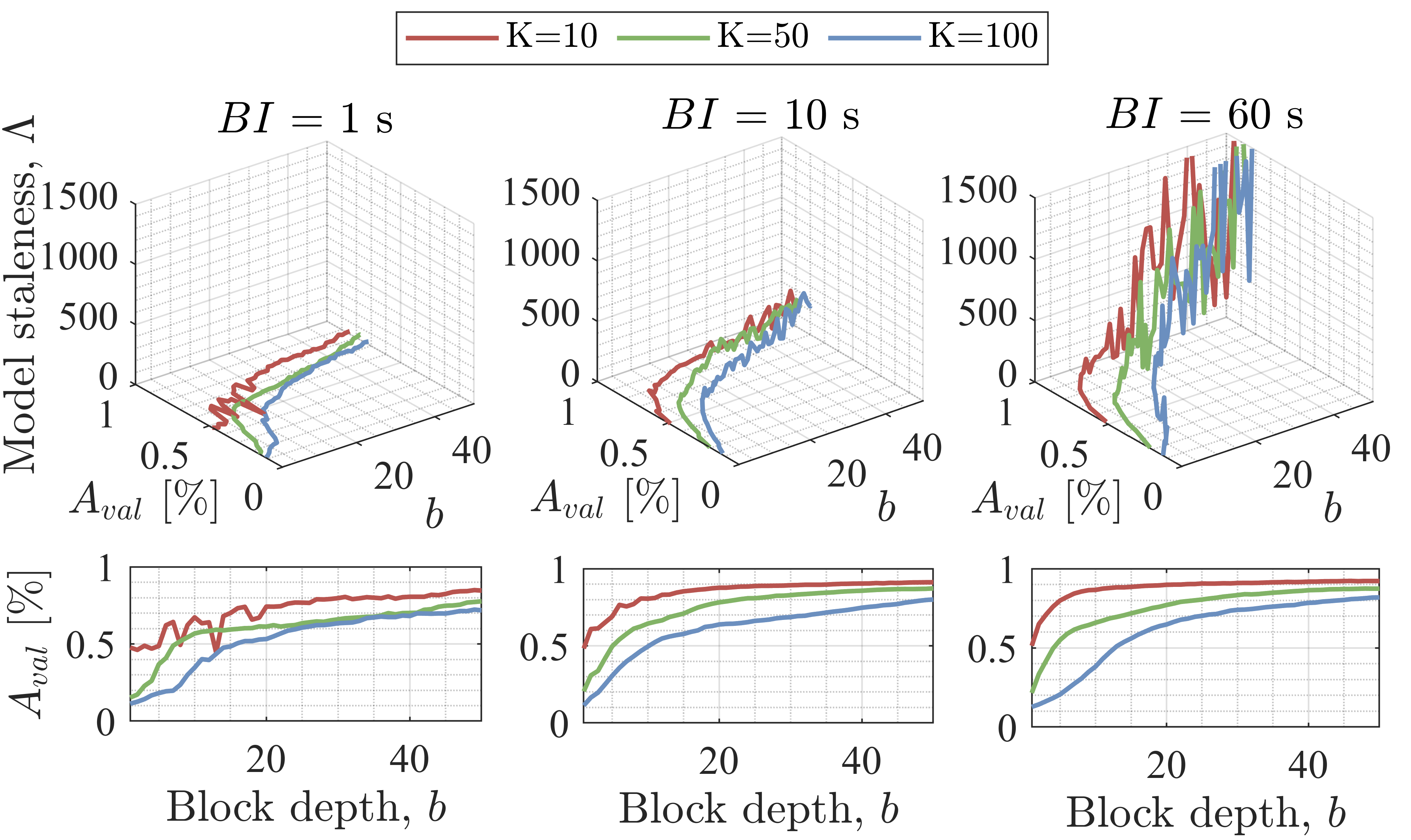}
\caption{\textcolor{black}{MNIST.}}
\label{fig:staleness_vs_accuracy_mnist}
\end{subfigure}
\begin{subfigure}{\columnwidth}
\includegraphics[width=\columnwidth]{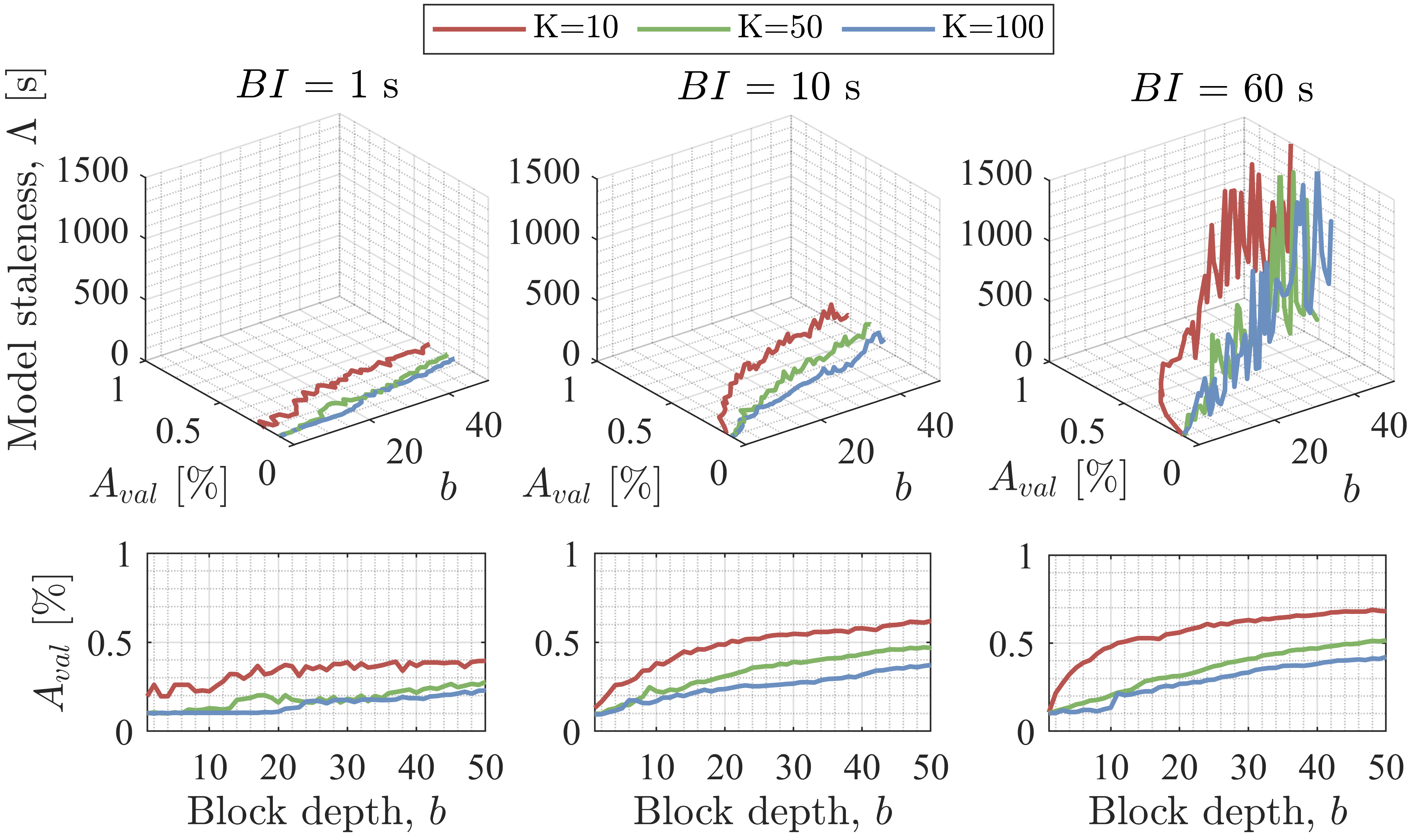}
\caption{CIFAR-10.}
\label{fig:staleness_vs_accuracy_cifar}
\end{subfigure}
\caption{Effect of model staleness on the validation accuracy: \textcolor{black}{a) MNIST, b) CIFAR-10}. The figures on the top show a 3D representation of the temporal evolution of model staleness and validation accuracy, while the plots on the bottom show the evolution of the validation accuracy in 2D.}
\label{fig:staleness_vs_accuracy_all}
\end{figure}

Next, we focus on model staleness and its impact on ML model accuracy. To that end, Fig.~\ref{fig:staleness_vs_accuracy_all} illustrates the temporal evolution of the model staleness and the validation accuracy experienced throughout the 50 first blocks of the main chain in each simulation. For the sake of reducing the impact of forks and focusing on \textcolor{black}{model} staleness, the results are shown only for $\text{C}_\text{link} = 100$~Mbps, where the fork probability is kept low in all the scenarios. Apart from that, the maximum number of local updates aggregated in a block, $\text{N}_\text{t}$, is fixed to 10. In the subplots on the top of Fig.~\ref{fig:staleness_vs_accuracy_all}, both \textcolor{black}{model} staleness and validation accuracy are displayed together in a 3D grid, while the corresponding 2D projection of the validation accuracy is plotted at the bottom of each figure for the sake of clarity.

\textcolor{black}{As shown in both Fig.~\ref{fig:staleness_vs_accuracy_mnist} and Fig.~\ref{fig:staleness_vs_accuracy_cifar},} model staleness remains stable for $BI=1$~s\textcolor{black}{, which indicates that the blockchain is able to process the model updates submitted by clients in time}. In contrast, \textcolor{black}{model staleness} increases very fast for each block when $BI=60$~s, as more and more local model updates remain unprocessed at the unconfirmed pool of transactions\textcolor{black}{, thus becoming outdated with respect to newer updates} (the blockchain cannot process all the local model updates). For $BI=1$~s, the achieved accuracy is significantly lower than in the other cases because FL devices are not fast enough to compute and provide local updates in time (as previously shown in Fig.~\ref{fig:time_evolution_val_accuracy_main_chain}), thus leading to \textcolor{black}{under-trained} models at each block depth. As for $BI=10$~s and $BI=60$~s, both solutions provide very similar accuracy, even if model staleness is much higher in the second case due to the slow pace at which the blockchain processes users' transactions (this is the opposite of what occurs with $BI=1$~s). This result suggests that model staleness is not detrimental to the model performance at all and, moreover, old local updates (i.e., updates computed from old global models) indeed contribute to sustaining the training of the federated models. Nevertheless, this conclusion is tied to the dataset used in this paper (CIFAR-10) and the distribution of the data over time, which in this case remains unchanged. However, other applications where data vary over time (which is associated with concept drift~\cite{tsymbal2004problem}) could be severely affected by staleness.

%% file: sections/6_future_directions.tex
\section{\textcolor{black}{Future Research Directions}}
\label{section:future_directions}

\subsection{\textcolor{black}{Blockchained FL optimization}}

\textcolor{black}{The analysis performed in Section~\ref{section:experimental_results} has demonstrated that the type of blockchain selected and its configuration have a high impact on the performance of the application running over it. In particular, model staleness and model inconsistencies, which are motivated by the mismatch between the application requirements and the blockchain performance, may result in significant performance degradation of FL if they are not kept under control. For that reason, blockchain optimization becomes particularly relevant for the FL use case. Whereas techniques like sharding or off-chain computation have been widely applied to improve current blockchains, other works have focused on optimizing blockchain configurations to better comply with the application requirements. In this regard, \cite{kim2019blockchained} focused on the block generation rate to minimize the end-to-end latency of blockchained FL applications, while \cite{wilhelmi2022end} targeted the optimization of the block size as a function of the users' activity.}

\textcolor{black}{Another important aspect for optimizing blockchained FL applications relates to the exchange of raw ML models between FL devices and blockchain miners. As shown in~\cite{guerra2023cost}, this approach entails significant communication and storage overheads, especially for complex ML models like VGG-16, whose size is over 500~MB~\cite{simonyan2014very}. To alleviate the communication burden, other distributed learning approaches such as Knowledge Distillation (KD)~\cite{zhu2021data}, much lighter than FL in terms of communication, might be compelling for its integration with blockchain.}

\subsection{\textcolor{black}{Trust and security enforcement through client selection}}

\textcolor{black}{The performance of blockchained FL applications can be affected by the quality and legitimacy of the contributions from the FL devices. To address these issues, we first find in FL client selection a prominent solution to speed up the performance of FL, thus relieving the burden on the blockchain. With FL client selection, user heterogeneity (e.g., in terms of computation and/or communication capabilities) can be addressed by selecting the best-performing nodes~\cite{nishio2019client}. Apart from FL client selection, blockchain inherent properties can be leveraged and extended to provide enhanced security and trust. Some prominent examples are enhanced authentication mechanisms~\cite{wang2021enabling} and trust evaluation and enforcement~\cite{wang2021heterogeneous}. Finally, there is the opportunity to adopt mechanisms in the blockchain for selecting client model updates to be mined, based on their degree of staleness (\textit{is the model update fresh enough?}).}

%% file: sections/7_conclusions.tex
\section{Conclusions}
\label{section:conclusions}

The blend of two disruptive technologies such as blockchain and AI can enable the proliferation of breakthrough innovations in the domain of collaborative computation. Blockchain and distributed ML solutions such as FL lead to a very profitable symbiosis in which trust and immutability are provided to decentralized applications, previously lacking security and privacy guarantees. In FL, blockchain allows getting rid of the figure of the central orchestrating server by replacing it with a democratic P2P network, thus contributing to relieving the issues of centralization (e.g., bottlenecking) and granting full control of the data to application participants. The partnership of blockchain and FL, however, poses a series of trade-offs that demand a joint design of the blockchain infrastructure and the underlying FL application. 

In this paper, we studied those trade-offs by analyzing the impact that different blockchain realizations have on FL performance in various scenarios. For that, we introduced an extension of BlockSim called  Block\textit{FL}sim---a blockchain simulator with embedded FL operations---to characterize FL applications running in a blockchain. Using  Block\textit{FL}sim, we studied model staleness and inconsistencies as a result of the blockchain operation and analyzed their impact on the FL model accuracy. Our results showed that model inconsistencies, resulting from fast blockchains with low communication capabilities, can largely contribute to lowering the FL model accuracy (up to 34\% less accuracy) when compared to more stable settings. When it comes to model staleness, we showed that its impact on the model accuracy is much lower than inconsistencies. Furthermore, we saw that, for the studied dataset, stale \textcolor{black}{model} updates can still contribute to improving the global model, which suggests that stale updates should not be discarded, so that already spent computational power is not wasted. \textcolor{black}{Future work includes the evaluation of staleness and model inconsistencies in experimental blockchain platforms, which would allow for providing further insights into the blockchain's decentralization-security-cost trilemma and the scalability of blockchained FL applications. In addition, the study of blockchain-native mechanisms for incentivizing the participation of FL devices, which in practice might be reluctant to invest computational power, is left as future work.}